# A 2D nanosphere array for atomic spectroscopy

**M. Romanelli[(1)], I. Maurin, P. Todorov[(2)], Chia-Hua Chan[(3)], D. Bloch**

*Laboratoire de Physique des Lasers, UMR 7538 du CNRS et de l'Université Paris 13, 99 avenue Jean-Baptiste Clément, 93430 Villetaneuse, France*
[(1)] *now at PALMS - UMR 6627 - Université de Rennes1-35042 Rennes cedex*
[(2)] *permanent address: Institute of Electronics, BAS, 72 Tzarigradsko Shosse boulevard, 1784 Sofia, Bulgaria*
[(3)] *permanent address: National Central University, Jung-Li City, Taoyuan, Taiwan 32054*

**Abstract.** We are interested in the spectroscopic behaviour of a gas confined in a micrometric or even nanometric volume. Such a situation could be encountered by the filling-up of a porous medium, such as a photonic crystal, with an atomic gas. Here, we discuss the first step of this program, with the generation and characterization of a self-organized 2D film of nanospheres of silica. We show that an optical characterization by laser light diffraction permits to extract some information on the array structure and represents an interesting complement to electron microscopy.

## 1. INTRODUCTION

Our group had extensively investigated the specific spectroscopic properties of thin [1] and very thin [2] vapour cells. For dilute vapour, the atomic free path restricted by the wall-to-wall distance, becomes anisotropic. This enhances the contribution of slow atoms, yielding a sub-Doppler signature in the spectroscopic response. Moreover, in the linear regime of absorption, one observes a coherent Dicke spectral narrowing [3] for a cell thickness $\lambda/2$ or below; it consists of a specific additive contribution of all velocity groups in the transient regime when the irradiation frequency is on the (Doppler-free) line-center. In the limit of short cell thickness (some tens of nanometres), it has even become possible to measure the van der Waals atom-surface interaction [4].

With such an exotic spectroscopy behaviour observed for a 1D- confined vapour, it is natural to wonder whether analogous effects can survive for a *two* or *three* dimensional confinement. Also, it is important to test the possibility of benefiting from the high resonant selectivity of gas media in the context of nanotechnologies and nanobjects. A 2D confinement has been achieved in recent experiments in gas-filled hollow-core photonic fibres with molecules and now atoms, and core size ~10 µm [5]. For a 3D confinement, we are searching ways to fill in a porous sample or photonic crystal with an atomic vapour. Here we focus on a possible 3D confinement by using a close-packed array of dielectric spheres of silica, very suitable to the use of standard resonant alkali vapour (no chemical reactivity up to 200 °C). We present here the first results that we have obtained in fabricating an ordered single layer of silica nanoparticles. Such an array could be sealed between two glass windows for future experiments (fig 1a).

## 2. NANOSPHERE ARRAY DEPOSITION AND CHARACTERIZATION

Our silica nanospheres are produced at National Central University, Taiwan. Their diameter is monodisperse and is chosen to be ~800 nm (the expected accuracy is better than 5 %). This sphere size is well suited for studying infrared transitions such as the resonance $D_1$ and $D_2$ lines of Cs, especially in view of a possible coherent Dicke effect when free paths are limited to less than $\lambda/2$. In order to obtain an ordered two-dimensional nanosphere array, we deposited a drop of a solution containing the nanoparticles onto a slightly tilted substrate (a simple microscope plate), previously treated with sulphuric acid and hydrogen peroxide ("piranha" solution) to make it clean and extremely hydrophilic [6]. Under properly controlled conditions, the particles self-assemble in an ordered 2-D film [7]. In our case, the best conditions turned out to be the followings: tilt angle of ~ 4°, and particle volume concentration of 0,4%. In order to slow down evaporation, the samples were enclosed in a small box (volume ~ 100 cm$^3$) during the phase of the array formation. An example of the obtained nanoparticles arrays is shown in fig. 1(b). In fig. 1(b), one distinguishes a large single ("crystalline") domain, whose typical size is of some tens of microns. The self-ordering of the array is limited by those particles with different size or shape, causing point defects and dislocations.

The size of spheres makes it possible to obtain sufficient insight by using an optical method, instead of scanning electron microscopy (SEM). The optical method has the advantage of being non-destructive (the SEM requires a metallization of the sample). Diffraction of a monochromatic plane wave by a regular array of identical scatterers is a well-known problem. Following [8], one easily obtains that the scattering angle θ (i.e. the angle between the scattered and the incoming wavevector) is given by [9]:

$$\sin\theta = \frac{2\lambda}{\sqrt{3}\,a} \sqrt{h^2 + k^2 - hk} \qquad (1)$$

where $\lambda$ is the laser light vacuum wavelength, a is the sphere diameter, and h and k are generic integers. It is only for a focused laser that the diffraction occurs on an (hexagonal) arrangement of closed-packed spheres, while, for a wider laser spot, many crystalline domains with different orientations are simultaneously irradiated, giving rise to a diffraction pattern that has the form of a ring. This is illustrated in fig. 1(c) and 1(d), where one sees respectively a hexagonal diffraction pattern produced by a He-Ne laser impinging at normal incidence, and diffraction by three crystalline domains with different orientations. From the largest beam waist that allows obtaining a single-domain pattern as in fig. 1(c), one estimates a single crystal domain size of about 60 x 60 µm². From eq. 1, the measured diffraction angle for the six first-order spots provides the value a= 812 ±20 nm for the lattice period. When measured with an UV laser beam, a is found to be 812 ±8 nm. By SEM we directly measure a = 816 ±8 nm (see fig.1(b)), in excellent agreement with the diffraction measurements. This demonstrates that optical diffraction allows measuring the sphere diameter with good precision. It is worth noting that the only relevant parameter is the distance between the scatterers. Therefore, the diffraction angle is the same for any spheres of the same size, regardless of their refractive index. Also, the efficiency of diffraction has been verified to depend on the polarization of the irradiating light, in a way compatible with the elementary predictions of Mie theory.

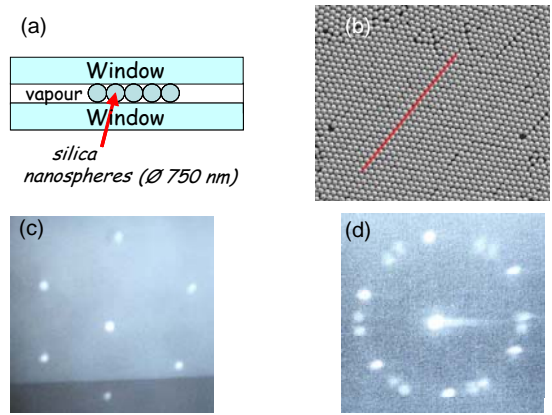

**Figure 1**. (a) Intended cell design for a system containing a single layer of silica spheres; (b) Image of the sample obtained by SEM. The length of the marker (*on line*: in red) is 20 µm. (c-d) Diffraction pattern produced by a He-Ne beam, on two different sites of the array.

## 3. DISCUSSION

We have obtained 2D nanoparticles arrays with a very simple method. The size of the single domains is satisfying for our future applications, since in a spectroscopic experiment we probe a sample region as large as the laser spot size, and in consequence we need the sample to behave as a single domain (photonic) crystal. A more serious issue is probably the sample thickness. We want to probe a gas inside the voids between the spheres. In order to do so, the array of nanospheres, presently deposited on a (microscope) glass plate, must be sealed between two plane windows, and must be in contact with the windows (if a space exists between the cell window and the film, the signal from the atoms in it will be dominant). So the thickness must be homogeneous over the whole film surface if the film has to be in close contact with the windows. The next step is to fabricate the vapour cell.